\documentclass[twocolumn,superscriptaddress,aps,prb]{revtex4-2}

\usepackage{graphicx}
\usepackage{dcolumn}
\usepackage{bm}
\usepackage{subfigure}
\usepackage{amsmath}
\usepackage{mathrsfs}
\usepackage{amssymb}
\usepackage{setspace}
\usepackage{array}
\usepackage{multirow}
\usepackage{float}
\usepackage{flushend}
\usepackage{footmisc}

\usepackage[pdfstartview=FitH,
CJKbookmarks=true,
colorlinks,
linkcolor=blue,
anchorcolor=blue,
citecolor=blue,
urlcolor=blue,
]{hyperref}

\usepackage{footmisc}

\renewcommand{\footnoterule}{
	\kern -4pt  
	\hrule width 0.18\linewidth height 0.6pt
	\kern 12pt 
}

\usepackage{footmisc}

\usepackage{braket}

\begin{document}
	
	\preprint{APS/123-QED}
	
	\title{Dissipation-Enhanced Localization in a Disorder-Free $\mathbb{Z}_2$ Lattice Gauge System}
	
	\author{Xuanpu Yang}
	\affiliation{School of Physics, Nankai University, Tianjin 300071, China}
	\affiliation{International Center for Quantum Materials and School of Physics, Peking University, Beijing 100871, China}

	\author{Xiang-Ping Jiang}
	\email{2015iopjxp@gmail.com}
	\affiliation{School of Physics, Hangzhou Normal University, Hangzhou, Zhejiang 311121, China}
	
	\author{Lei Pan}%
	\email{panlei@nankai.edu.cn}
	
	\affiliation{School of Physics, Nankai University, Tianjin 300071, China}

	\begin{abstract} 
The $\mathbb{Z}_2$ lattice gauge model, as the simplest realization of a lattice gauge theory, exhibits rich and unconventional physics. One of its most remarkable features is disorder-free localization, where localization emerges not from explicit quenched disorder but from static background $\mathbb{Z}_2$ gauge charges, leading to persistent memory of the initial state. In this work, we investigate the dissipative dynamics of the $\mathbb{Z}_2$ lattice gauge model by coupling it to a Markovian environment. We find that quantum dissipation can enhance localization: memory of the initial state is retained more robustly under dissipative evolution than under unitary dynamics. This dissipation-induced enhancement of localization persists across a variety of initial states, indicating that the effect is not limited to fine-tuned configurations. Our results demonstrate that dissipation, often associated with decoherence and thermalization, can in fact serve as a powerful tool for stabilizing non-ergodic behavior in gauge-constrained quantum systems.

	\end{abstract}
	
	
	\maketitle
	\section{Introduction}

Gauge theories play a central role in both high-energy physics and condensed matter systems, providing a powerful framework to describe fundamental interactions and emergent phenomena \cite{Kogut_RMP1979}. In recent years, lattice gauge theories have attracted growing attention as platforms for exploring exotic quantum phases and non-equilibrium dynamics in controlled quantum simulators~\cite{Banerjee_PRL2012,Zohar_RPP2016,Martinez_Nature2016,Gauge_exp1,Gauge_exp2,Gauge_exp3,Gauge_exp4,Gauge_exp5,Gauge_exp6,Gauge_exp7,Gauge_exp8}. Among them, the $\mathbb{Z}_2$ lattice gauge model stands out as the simplest nontrivial realization of gauge theory on a lattice, while still capturing a variety of intriguing phenomena~\cite{Z2_theory1,Z2_theory2}
One particularly remarkable feature of the $\mathbb{Z}_2$ lattice gauge model is the so-called disorder-free localization. Unlike conventional localization that arises from explicit quenched disorder~\cite{Anderson_PR1958,Nandkishore_ARCMP2015,AndersonTrans2008}, localization in this system emerges from static configurations of background $\mathbb{Z}_2$ charges, which effectively induce a sector-dependent random potential~\cite{Z2_theory1,Z2_theory2}. This emergent disorder fragments the Hilbert space into dynamically disconnected sectors, each exhibiting robust retention of the initial state’s memory during unitary evolution.

While localization in closed quantum systems has been extensively studied, the role of dissipation in such gauge-constrained, disorder-free systems remains largely unexplored. The theory of open quantum systems has a long history and has recently regained attention owing to advances in controlling both dissipation and Hamiltonian parameters~\cite{Exp1,Exp2,Exp3,Exp4,Exp5,Exp6,Exp7,Exp8,Exp9,Exp10,Exp11,Exp12,Exp13}. These developments have enabled studies of interacting open systems~\cite{OpenMB0,OpenMB1,OpenMB2,OpenMB3,OpenMB4,OpenMB5,OpenMB6,OpenMB7,OpenMB8,OpenMB9,OpenMB10,OpenMB11,OpenMB12,OpenMB13,OpenMB14,OpenMB15,OpenMB16,OpenMB17}, as well as systems subject to disorder or quasiperiodic potentials~\cite{OpenMBL1,OpenMBL2,OpenMBL3,OpenMBL4,OpenDisorder1,OpenDisorder2,OpenDisorder3,OpenDisorder4,OpenDisorder5,OpenDisorder6,OpenDisorder7,OpenDisorder8,OpenDisorder9,OpenDisorder10,OpenDisorder11,OpenDisorder12,OpenDisorder13,OpenDisorder14}, and non-Hermitian lattice gauge theory \cite{NH_Z2}.
Conventionally, quantum dissipation is associated with decoherence, thermalization, and the loss of information about initial conditions. Hence, dissipation is typically expected to destroy quantum coherence and suppress localization. Indeed, dephasing from environmental coupling often leads to delocalized steady states and enhanced transport~\cite{OpenDisorder_transport1,OpenDisorder_transport2,OpenDisorder_transport3}. 
However, recent studies have shown that properly engineered dissipation can stabilize localized phases~\cite{Yusipov17} and can induce localization-delocalization transitions in disordered and quasiperiodic systems~\cite{WYC_PRL,Jiang_3D}. Moreover, specially designed dissipation can drive systems into non-thermal steady states, including many-body localized phases~\cite{Yusipov18,WYC_MBL} and quantum many-body scars~\cite{Diss_scar1,Diss_scar2,Diss_scar3}, and protect certain features of non-ergodic dynamics.
These advances naturally raise the question of how dissipation influences disorder-free localization in lattice gauge systems and what physical consequences it may entail.

%

In this work, we investigate how dissipation influences localization in the $\mathbb{Z}_2$ lattice gauge model by coupling the system to a Markovian bath. Remarkably, we find that a class of jump operators, acting pairwise on fermions, can enhance the localization behavior: the quantum fidelity with respect to the initial state decays more slowly under dissipative dynamics than in the unitary case. This dissipation-enhanced memory retention signifies a counterintuitive role of quantum dissipation in stabilizing non-ergodic behavior.
Furthermore, the influence of dissipation phase on the steady-state behavior reveals that the degree of localization can be finely tuned through phase control. Finally, by adjusting the jump range in the dissipation operators, we demonstrate that enhanced localization persists across different symmetry sectors.
Our findings reveal an unexpected regime in which dissipation, instead of degrading coherence, can be exploited to reinforce localization and memory in disorder-free quantum systems. This opens a pathway to stabilize memory and engineer non-ergodic steady states through tailored dissipation, underscoring the interplay of gauge constraints, non-equilibrium dynamics, and reservoir design with relevance to quantum memory and thermalization control.

\section{$\mathbb{Z}_2$ Lattice Gauge Model and the emergent disorder }
\label{Property}
	
In this section, we briefly introduce the model to be studied, namely the $\mathbb{Z}_2$ lattice gauge model and associated emergent disorder. 
Consider a one-dimensional chain of spinless fermions $\hat{f}_i$ residing on the lattice sites, coupled to $\mathbb{Z}_2$ bond spins (gauge fields) $\hat{\sigma}_{i,i+1}^{x,z}$ defined on the links, the system Hamiltonian is given by 

\begin{align}
	\hat{H} = -J \sum_{\langle ij \rangle} \hat{\sigma}_{ij}^z \hat{f}_j^\dagger \hat{f}_k
	- h \sum_j \hat{\sigma}_{j-1,j}^x \hat{\sigma}_{j,j+1}^x. \label{Ham1}
\end{align}
Here $\langle ij \rangle$ denotes nearest-neighbor sites and $\hat{f}_j$ ($\hat{f}_{j}^\dagger$) is the annihilation (creation) operator of the spinless fermion at site $j$, and $\hat{\sigma}_{i,i+1}^{x}$ ($\hat{\sigma}_{j,j+1}^{z}$) denote Pauli operators acting on the link between sites $j$ and $j+1$. They satisfy the usual algebra $(\hat{\sigma}_{j,j+1}^\alpha)^2=1$,
 $\left\{\hat{\sigma}_{j,j+1}^x,\hat{\sigma}_{j,j+1}^z\right\}=0$, and $\left[\hat{\sigma}_{j,j+1}^z,\hat{\sigma}_{j,j+1}^x\right]=2i\hat{\sigma}_{j,j+1}^y$. The first term of the Hamiltonian \eqref{Ham1} describes nearest-neighbor hopping of fermions with strength $J$, accompanied by a $\mathbb{Z}_2$ gauge field on the corresponding bond. The second term represents the dynamics of gauge fields with strength $h$, and being noncommutative with $\hat{\sigma}_{j,j+1}^z$, provides kinetic fluctuations of the $\mathbb{Z}_2$ degrees of freedom. One can find the Hamiltonian contains a large set of conserved charges defined as $\hat{q}_j = (-1)^{\hat{n}_j}\hat{\sigma}_{j-1,j}^x \hat{\sigma}_{j,j+1}^x$, where $\hat{n}_j=\hat{f}_j^\dagger \hat{f}_j$ is the fermion number operator. Each $\hat{q}_j$ has eigenvalues $\pm 1$, and they mutually commute as well as commute with the Hamiltonian, i.e., $\left[\hat{H},\hat{q}_j\right]=0$ and $\left[\hat{q}_i,\hat{q}_j\right]=0$, reflecting the exact $\mathbb{Z}_2$ gauge invariance of the model. These conserved quantities generate local $\mathbb{Z}_2$ gauge symmetries under which the Hamiltonian remains invariant. Concretely, the transformations are implemented through the unitary operator
$\hat{U}\!\left(\{\theta_i\}\right) = \prod_j \hat{q}_j^{(1-\theta_j)/2}$, with $\quad \theta_j=\pm 1$, which act on the fields as  
\begin{align}
\hat{f}_j \mapsto \theta_j \hat{f}_j, \quad  \hat{f}_{j}^\dagger \to \theta_i \hat{f}_{j}^\dagger, \quad
\hat{\sigma}_{j,j+1}^z \mapsto \theta_j \hat{\sigma}_{j,j+1}^z \theta_{j+1} .
\end{align}

Although the Hamiltonian contains no explicit disorder, an effective randomness emerges once the dynamics are expressed in terms of the $\mathbb{Z}_2$ charges. To show this, we begin with a duality transformation \cite{duality1,duality2} of the $\sigma$ operators, by introducing spin-$1/2$ variables $\tau$ defined on the lattice sites
\begin{align}
\hat{\tau}_j^z =\hat{\sigma}_{j-1,j}^x\hat{\sigma}_{j,j+1}^x, \qquad \hat{\tau}_j^x \hat{\tau}_{j+1}^x = \hat{\sigma}_{j,j+1}^z , 
\end{align}
and the local charge operator becomes
\begin{align}
	\hat{q}_j = (-1)^{\hat{n}_j}\hat{\tau}_j^z.
\end{align}
Defining a composite fermion operator $\hat{c}_j = \hat{\tau}_j^x \hat{f}_j$, which binds the spinless fermion to the dual spin and then the Hamiltonian is expressed in terms of spinless fermions $\hat{c}_j$ and conserved charges $\hat{q}_{j}$ 
\begin{align}
	 	\hat{H}=-J\sum_{\langle jk\rangle}\hat{c}_{j}^{\dagger}\hat{c}_{k}+2h\sum_{j}\hat{q}_{j}(\hat{c}_{j}^{\dagger}\hat{c}_{j}-1/2). \label{Ham_Z2}
\end{align}
Here we have used the identity $\hat{n}_j = \hat{f}_j^\dagger \hat{f}_j = \hat{c}_j^\dagger \hat{c}_j$, together with $(\hat{\tau}_j^x)^2=1$. The canonical anticommutation relation $\{ \hat{c}_j, \hat{c}_k^\dagger \} = \delta_{jk}$ follows straightforwardly. 
In this representation, the first term describes a simple tight-binding model for the composite fermions, while the second term acts as an on-site potential determined by the eigenvalues $q_j = \pm 1$ of $\hat{q}_j$. Each configuration of ${q_j}$ defines a distinct gauge sector, effectively mapping the Hamiltonian onto a free-fermion problem with a pattern of binary on-site energies serving as a disorder potential. Consequently, the Hilbert space decomposes into independent sectors labeled by ${q_j}$, and the disorder in the dynamics arises from the distribution of these $\mathbb{Z}_2$ charges.
  	 
%
	
The presence of emergent disorder gives rise to localization phenomena that protect information about the initial state. This effect, often referred to as \emph{disorder-free localization}, originates from the binary $\mathbb{Z}_2$ charge configuration acting as a source of effective on-site disorder. In contrast to conventional Anderson localization, where randomness is introduced externally through quenched disorder in the Hamiltonian parameters, here the disorder is dynamical and arises intrinsically from the conserved $\mathbb{Z}_2$ gauge charges. Each site is subject to a local binary field determined by its charge configuration, and the ensemble of such fields effectively mimics a disordered potential landscape for the itinerant fermions. The relative strength of the emergent disorder is characterized by the ratio $h/J$, where $h$ controls the coupling to the conserved charges and $J$ denotes the hopping amplitude. Larger values of $h/J$ correspond to a stronger binary potential landscape, thereby reducing the ability of the fermions to delocalize across the lattice and enhancing the localization effect. In this sense, tuning $h/J$ provides a direct handle for interpolating between a delocalized regime, where thermalization dominates, and a localized regime, where memory of the initial state is retained for long times. To quantitatively probe how much the initial configuration's memory is preserved, we evaluate the fidelity between the instantaneous state $\rho(t)$ and the initial state $\rho_0$, defined as \cite{fidelity} 
\begin{align} 
	\begin{split} F(\rho,\rho_0) = \text{tr}\Big(\sqrt{(\rho)^{\frac{1}{2}}\rho_0(\rho)^{\frac{1}{2}}}~\Big). \end{split}
 \end{align} 
 
\begin{figure}[!ht] 
\includegraphics[width=1.0\linewidth]{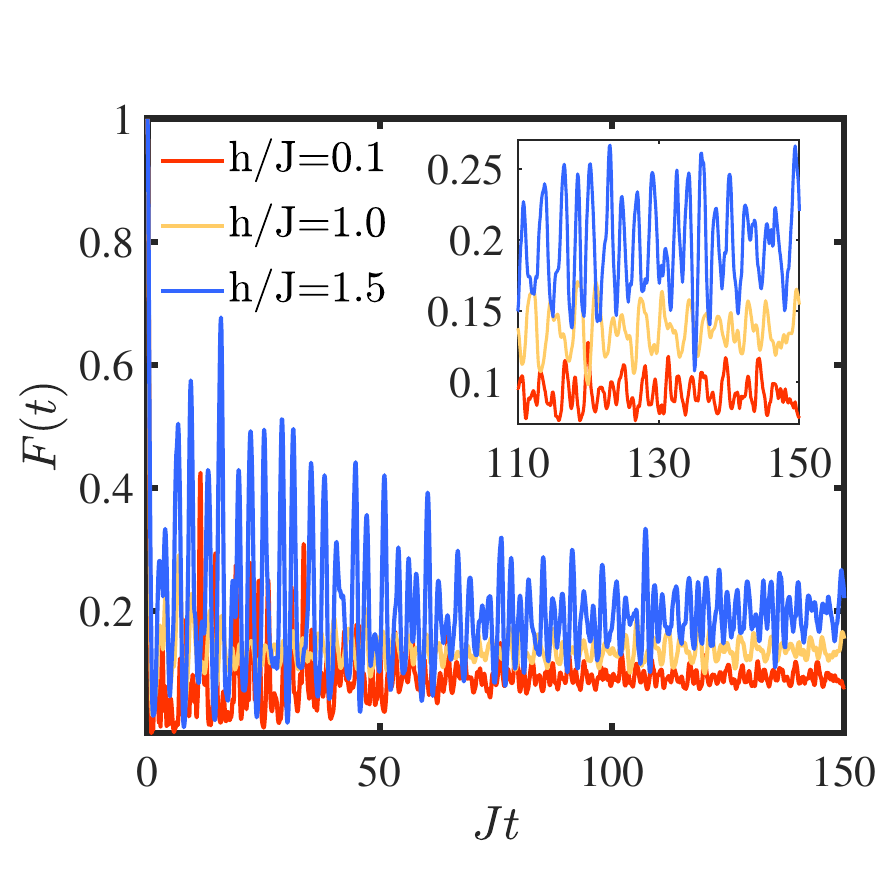} 
\caption{Unitary time evolution of the fidelity $F(t)$ with respect to the initial state for several representative values of the emergent disorder strength $h/J$ in a system of length $L=10$. The main panel shows the global behavior of $F(t)$, highlighting how increasing $h/J$ enhances localization and suppresses thermalization, thereby improving memory retention of the initial state. The inset provides a magnified view of the late-time regime, restricted to the interval $110 \leq t \leq 150$, where the differences between the curves become more visible. In this zoomed region, larger $h/J$ values are seen to stabilize higher fidelity plateaus, further illustrating the role of emergent disorder in preserving quantum information. The system size is $L=10$ and the initial state is chosen as a charge density wave state $\ket{Z_2}=\ket{1010101010}$.} \label{fig1}
\end{figure}

 A higher fidelity reflects a stronger overlap with the original state and therefore more robust memory preservation, whereas a rapid decay of $F(t)$ signals thermalization and loss of information. Figure~\ref{fig1} illustrates the unitary dynamics of the fidelity for three representative values of $h/J$ in a chain of length $L=10$. The initial state is chosen as a ${Z}_2$ state $\ket{Z_2}=\ket{1010\dots 1010}$, where ``$1$’’ denotes an occupied fermion and ``$0$’’ denotes an empty one. This state possesses a $\mathbb{Z}_2$ symmetry, that is, it is invariant under a two-site translation. As plotted in the Fig.\ref{fig1}, although the $F(t)$ decreases from its initial value of $1$, it saturates to a finite level at long times for any $h \neq 0$. Moreover, the asymptotic limit $F(t \rightarrow \infty)$ grows monotonically with $h/J$, as illustrated in the inset. This behavior indicates that the system retains memory of its initial configuration, a consequence of ergodicity breaking induced by emergent disorder. This behavior clearly demonstrates that emergent disorder serves as a mechanism for stabilizing localization and retaining quantum memory, even in the absence of external quenched randomness. 

\section{Lindblad Dynamics of $\mathbb{Z}_2$ Lattice Gauge Model}
\label{Model}

In this section, we focus on how dissipation influences the emergent localization in the dynamical evolution of the $\mathbb{Z}_2$ lattice gauge system. Suppose the system interacts with an external reservoir, described by the total Hamiltonian 
\begin{align}
	\hat{H}_T = \hat{H}_S+\hat{H}_R+\hat{H}_{SR},
\end{align}
where $\hat{H}_S$ and $\hat{H}_R$ denote Hamiltonians of the system and reservoir, while $\hat{H}_{SR}$ describes the coupling between them. 
Under the Born-Markov approximation~\cite{Moy1999,Breuer2002}, tracing out the reservoir's degrees of freedom brings about the Lindblad master equation~\cite{Lindblad1,Lindblad2}
\begin{align} \frac{d\rho(t)}{dt} = \mathscr{L}[\rho(t)] = -i[H_{S}, \rho(t)] + \mathcal{D}[\rho(t)], \label{Lindblad_Eq}
\end{align} 
where $\rho(t)$ is the reduced density matrix of the system, and $\mathscr{L}$ denotes the Liouvillian superoperator, which guarantees both trace preservation and complete positivity. The first commutator term generates coherent unitary dynamics, while the dissipator $\mathcal{D}[\rho(t)]$ encodes the effects of the environment
\begin{align}
	\mathcal{D}[\rho(t)]= \sum_{j}\Gamma_j \Big( \hat{O}_j \rho \hat{O}_j^\dagger - \tfrac{1}{2}\big\{\hat{O}_j^\dagger \hat{O}_j,\rho\big\}\Big),
\end{align}
Here $\{\cdot,\cdot\}$ stands for the anticommutator, $\hat{O}_{j}$ is the quantum jump (dissipative) operator associated with lattice site $j$, and $\Gamma_{j}$ specifies the corresponding dissipation strength. Just as the Hamiltonian governs the Schr{\"o}dinger equation of isolated systems, the Liouvillian superoperator determines the full non-equilibrium dynamic behavior of open quantum systems. The formal solution of Eq.~\eqref{Lindblad_Eq} takes the form $\rho(t) = e^{\mathscr{L}t}[\rho_{0}]$, and in the asymptotic limit $t \to \infty$, the system converges to a stationary state $\rho_{\mathrm{ss}}$, which corresponds to the right eigenvector of $\mathscr{L}$ with eigenvalue zero.

In the presence of dissipation, the system eventually relaxes to a steady state that is independent on the choice of initial states. As discussed above, in purely unitary dynamics, localization preserves part of the memory of the initial state, resulting in a finite long-time quantum fidelity. Consequently, the steady-state fidelity obtained from the Lindblad dynamics provides a diagnostic of localization under dissipation: an increased (decreased) fidelity indicates that dissipation enhances (suppresses) localization.

In conventional scenarios, dissipation quickly erases any memory of initial states. For instance, choosing the on-site dephasing $\hat{O}_j=\hat{n}_j$, the system will be driven into an infinite-temperature state or maximally mixed state, thereby all information about the initial configuration is removed. In this sense, dissipation destroys localization, which is also the widely accepted expectation.
However, we find that under certain types of dissipative operators, the signatures of localization are not weakened but instead reinforced, as reflected in the enhanced fidelity with respect to the initial state.
The specific class of dissipation operators takes the following form
\begin{align}
	\hat{O}_{j} = \left( \hat{c}_{j}^\dagger + e^{i\alpha}\hat{c}_{j+l}^\dagger \right)
	\left( \hat{c}_{j} - e^{i\alpha}\hat{c}_{j+l} \right),
	\label{O_j_alpha}
\end{align}
where $\alpha$ is the dissipation phase and each operator acts on a pair of sites separated by a distance $l$. 
This type of dissipation operators were first introduced in earlier studies~\cite{Jump1,Jump2}. One implementation relies on a one-dimensional Bose-Hubbard chain~\cite{BHchain}, while another scheme based on Raman optical lattices allows the dissipation phase to be tuned between $0$ and $\pi$~\cite{WYC_PRL}.
From a physical perspective, each operator acts simultaneously on two sites, $j$ and $j+l$, and alters the relative phase between them. For example, with dissipation phase $\alpha=0$, the operator takes the form $\hat{O}_j = (\hat{c}_{j}^\dagger + \hat{c}_{j+l}^\dagger)(\hat{c}_{j} - \hat{c}_{j+l})$, which converts an anti-symmetric mode into a symmetric mode. In contrast, when $\alpha=\pi$, the operator becomes $\hat{O}_j = (\hat{c}_{j}^\dagger - \hat{c}_{j+l}^\dagger)(\hat{c}_{j} + \hat{c}_{j+l})$, removing the symmetric mode shared by the two sites and generating an anti-symmetric excitation.
Next, we investigate the impact of a specific class of quantum dissipation on the $Z_2$ lattice gauge system. We demonstrate that, rather than erasing memory of the initial configuration, certain dissipative processes can in fact reinforce it, leading to enhanced memory retention. Specifically, we study the $Z_2$ lattice gauge model \eqref{Ham_Z2} with system size $N=10$ and coupling ratio $h/J=0.5$. The initial state is chosen to be $\ket{Z_2}=|1010101010\rangle$. In addition, there exists another state with the same symmetry, obtained by translating all particles one site to the right in real space, yielding the state $\ket{Z_2^{\prime}}=|0101010101\rangle$.

\begin{figure}[!ht]
	\includegraphics[width=1.0\linewidth]{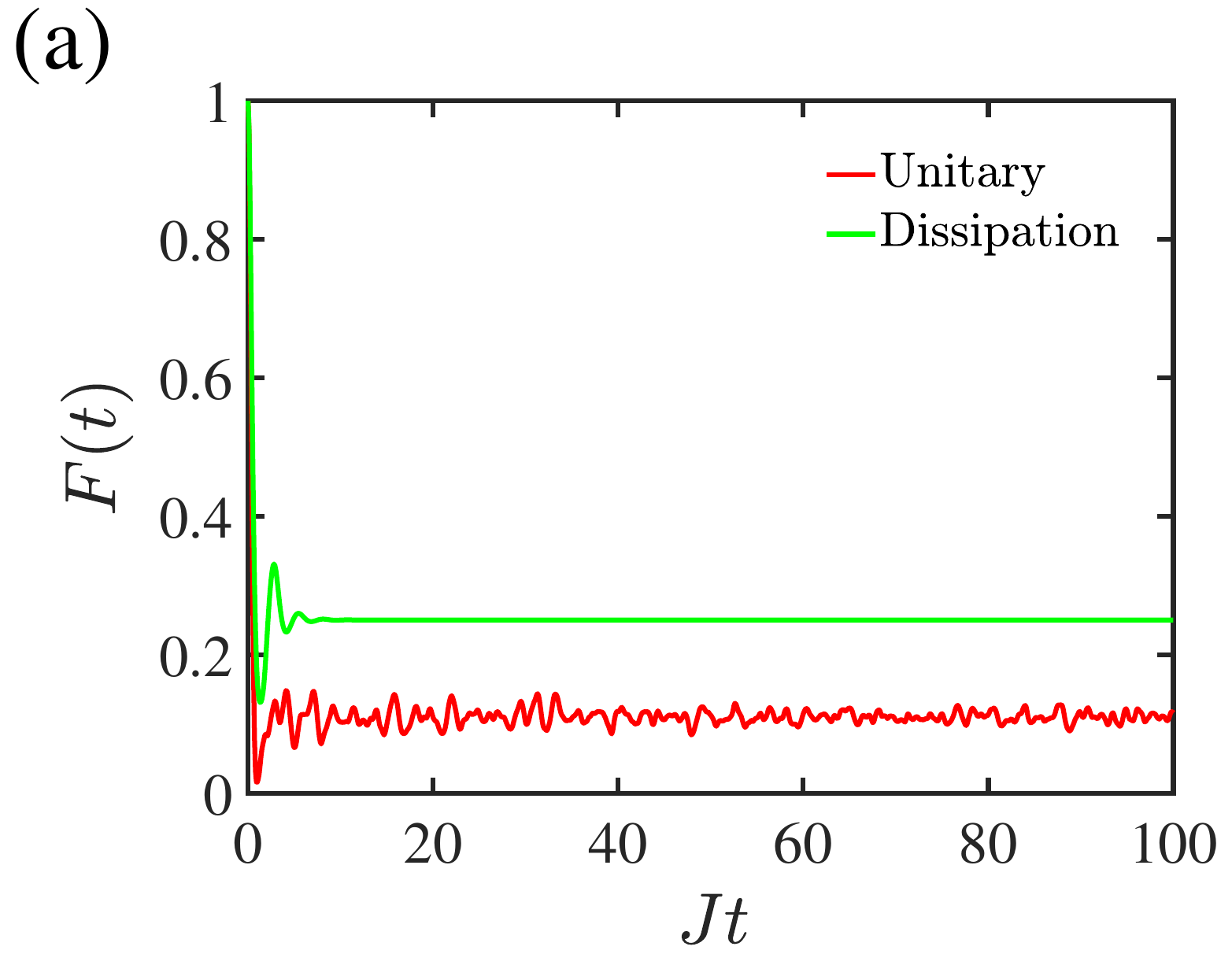}
	\includegraphics[width=1.0\linewidth]{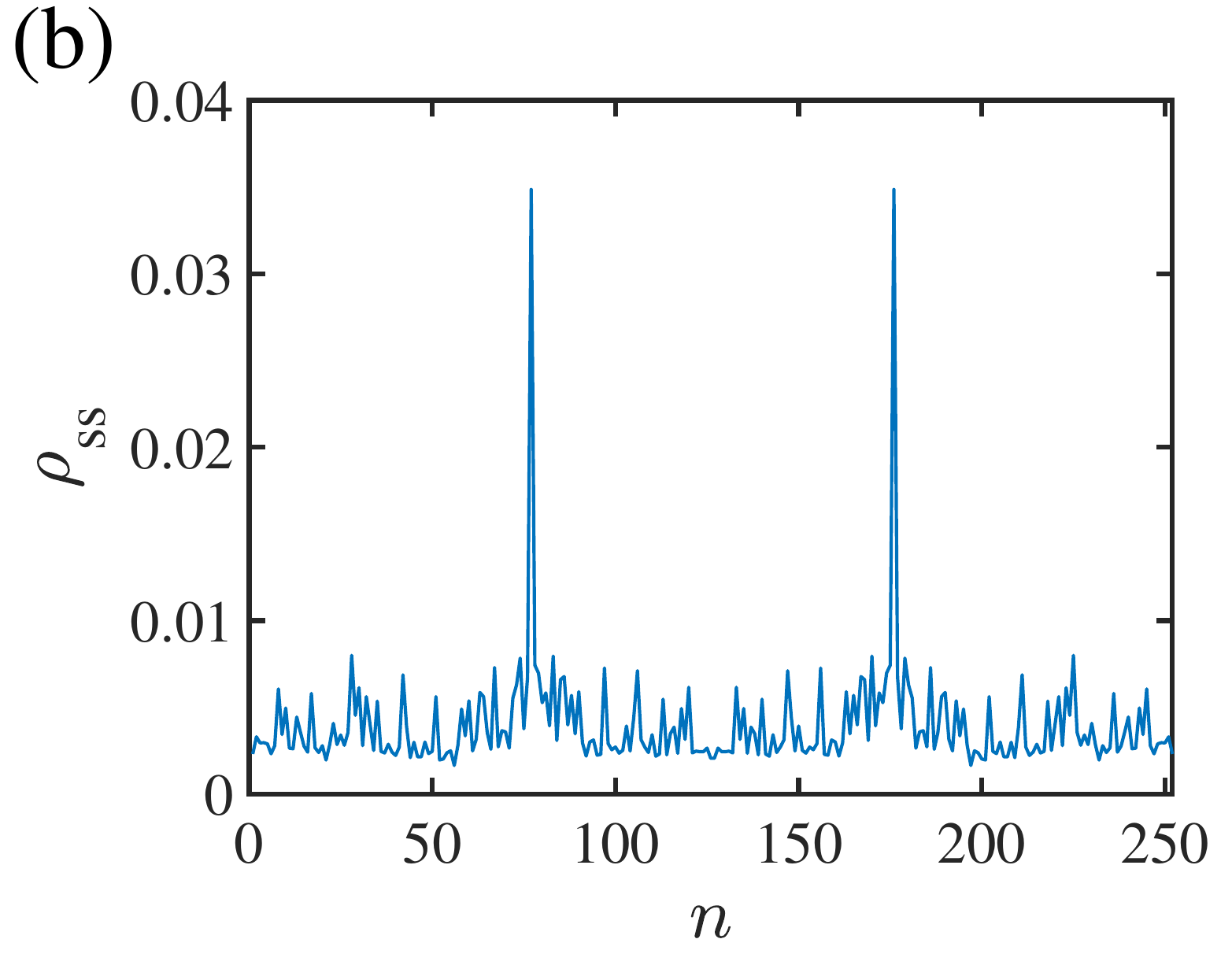}
	\caption{(a) Fidelity with respect to the initial state $\ket{Z_2}$ under unitary (red) and dissipative (green) evolution for $L=10$ and $h/J=0.5$. Dissipation leads to enhanced memory retention compared to the purely unitary dynamics. (b) Diagonal elements of the steady-state density matrix in the local basis. The presence of two sharp peaks indicates that the steady state predominantly resides in $\ket{Z_2}=\ket{1010101010}$ and $\ket{Z'_2}=\ket{0101010101}$.}
	\label{Fig2}
\end{figure}

In what follows, we analyze both the unitary dynamics governed by the Liouville equation and the open-system evolution described by the Lindblad master equation. For the latter, we set the dissipation strength to $\Gamma = 1$, choose a jump amplitude $l=2$, and fix the dissipation phase at $\alpha = 0$. Under these conditions, the dissipation is represented as $O_{j} = (c_{\boldsymbol{j}}^\dagger + c_{\boldsymbol{j}+2}^\dagger)(c_{\boldsymbol{j}} - c_{\boldsymbol{j}+2}).$
We set $h/J=0.5$ and $\Gamma=1$, and compare the dynamics with and without dissipation, starting from initial state $\ket{Z_2}$ . The results are shown in Fig.~\ref{Fig2}(a). Remarkably, the dissipative evolution (green) preserves memory of the state $\ket{Z_2}$ more effectively than the purely unitary evolution (red).
%
For convenience, we order the local basis states according to the magnitude of their corresponding binary numbers. For example, the system with length $L=10$ in the half-filled condition, i.e., $N_f=L/2=5$, the Hilbert space dimension is $252$. Each possible state is assigned an index: for example, $|0000011111\rangle$ corresponds to 1st state, and  $|0000101111\rangle$ to 2nd state, and so on, up to $|1111100000\rangle$, which is labeled as state $252$. Within this convention, the state $\ket{Z_2}=|1010101010\rangle$ corresponds to No. $176$ basis, while its translation state $\ket{Z_2^{\prime}}=|0101010101\rangle$ corresponds to No. $77$ basis.

\begin{figure}[!ht]
	\includegraphics[width=1.0\linewidth]{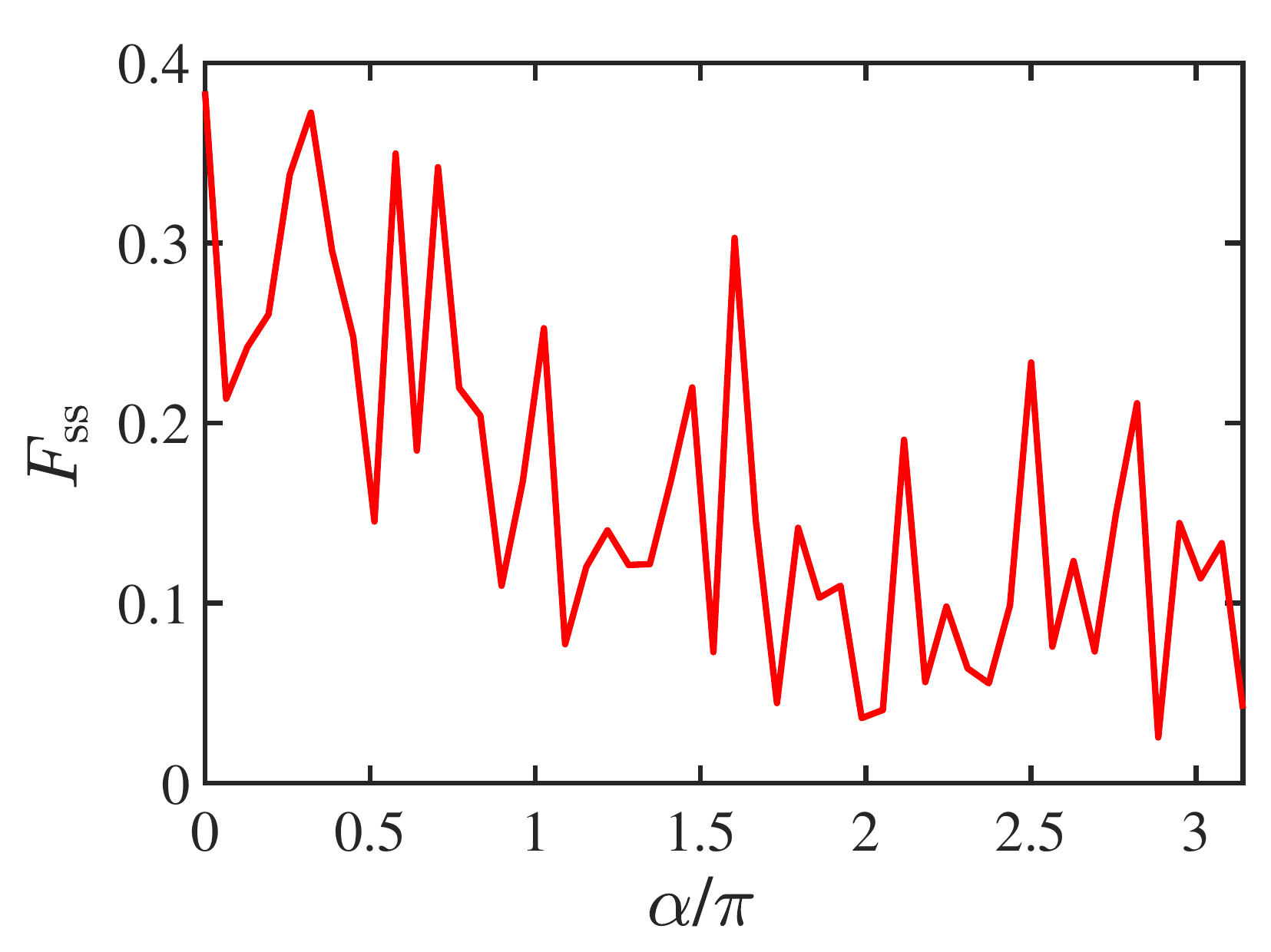}
	\caption{Steady-state fidelity $F_{ss}$ as a function of the dissipation phase $\alpha$ for $L=10$ and $\Gamma/J=0.5$. he fidelity decreases as $\alpha$ increases, exhibiting an overall downward trend accompanied by pronounced oscillations, which signals the reduced robustness of the  memory for $Z_{2}$ state.}
	\label{Fig3}
\end{figure}

To further elucidate this effect, Fig.~\ref{Fig2}(b) displays diagonal elements of the steady-state density matrix in the local basis. One can see there are sharp peaks which correspond to ${Z_2}$ and ${Z_2^{\prime}}$ state respectively. This explains why the fidelity of the $Z_2$ initial state is enhanced under this form of dissipation.
We further investigate the influence of the dissipation phase on the steady-state fidelity. To this end, Fig.~\ref{Fig3} plots the steady-state fidelity $F_{ss}$ for the $Z_2$ state as a function of the dissipation phase $\alpha$. It can be observed that, overall, the fidelity decreases (with oscillations) as$\alpha$ increases, while exhibiting relatively large peaks at multiple phase values. This indicates that the enhanced localization effect is most robust at $\alpha=0$ and also manifest at multiple dissipative phases.


\section{Dissipation Behaviors of other initial state}
\label{dissipation_localization}
In the previous section, we used the initial state $\ket{Z_2}$ as an example to demonstrate how dissipation can enhance disorder-free localization. A natural question is what about other types of initial states? Can the dissipation mechanism we propose also preserve memory more effectively than the unitary evolution? In this section, we turn to other initial configurations and show that, by tuning the parameters of the dissipative operators, the system's memory of initial states can likewise be reinforced. 

To address this, we first consider the case where the initial state is a $\mathbb{Z}_3$ symmetric state $\ket{Z_3}=\ket{100100\cdots100100}$. As in the previous discussion, we compute both the unitary dynamics without dissipation and the dissipative dynamics. 
By comparing the long-time quantum fidelity in the two scenarios, we determine whether the disorder-free localization effect is enhanced. In this case, the form of dissipation operators \eqref{O_j_alpha} remains unchanged, but the jump distance $l$ should be chosen to match the underlying symmetry of the initial state. Here we take $l=3$ to match the jump range, and the total lattice size $L$ is taken as an integer multiple of $l$ which also ensures compatibility with the translation symmetry of the $\ket{Z_{3}}$ state. The corresponding dissipation operator is therefore given by
\begin{align}
	\hat{O}_{j} = \left( \hat{c}_{j}^\dagger + \hat{c}_{j+3}^\dagger \right)\left( \hat{c}_{j} - \hat{c}_{j+3} \right).
\end{align}

\begin{figure}[!t] 
	\includegraphics[width=1.0\linewidth]{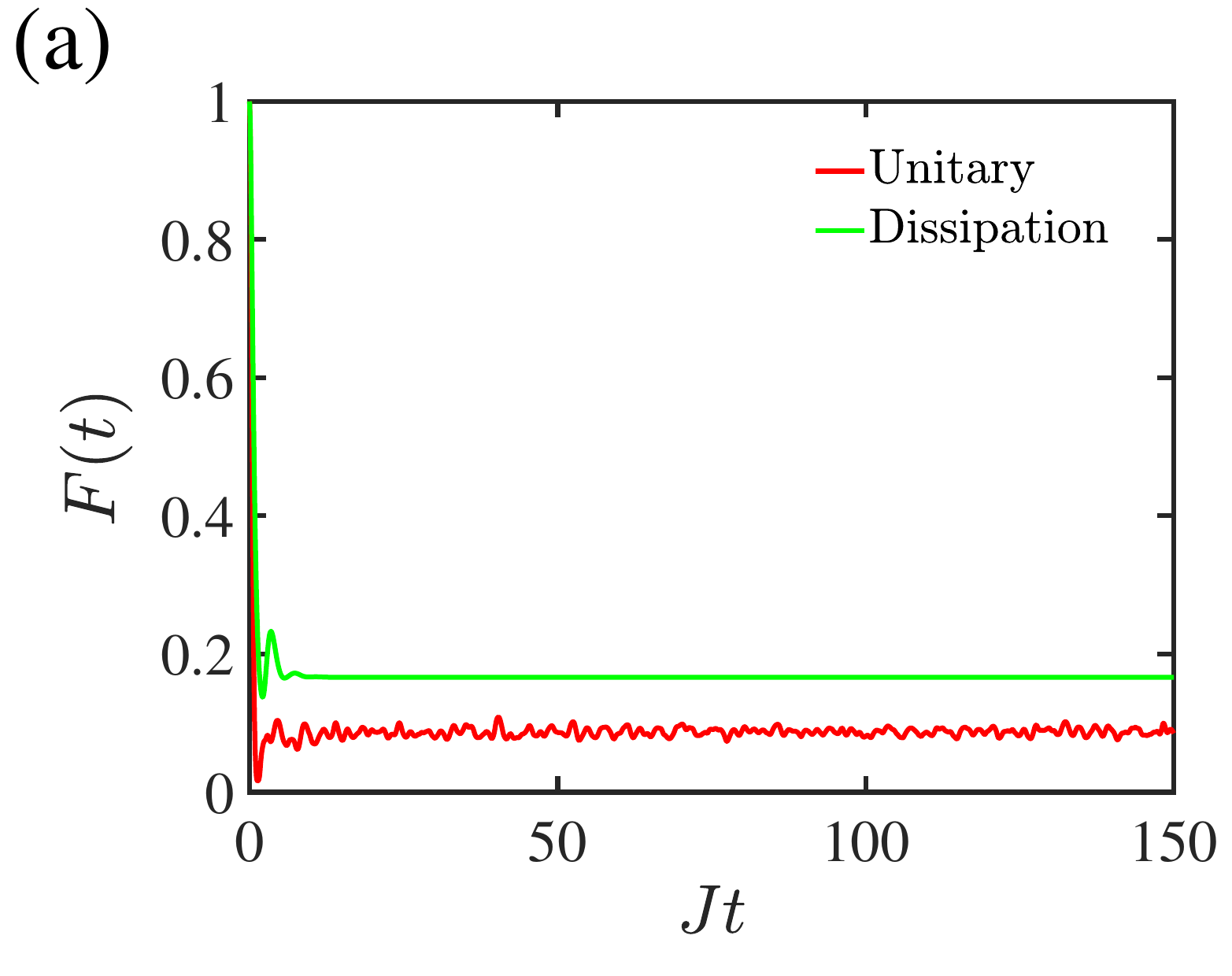} 
	\includegraphics[width=1.0\linewidth]{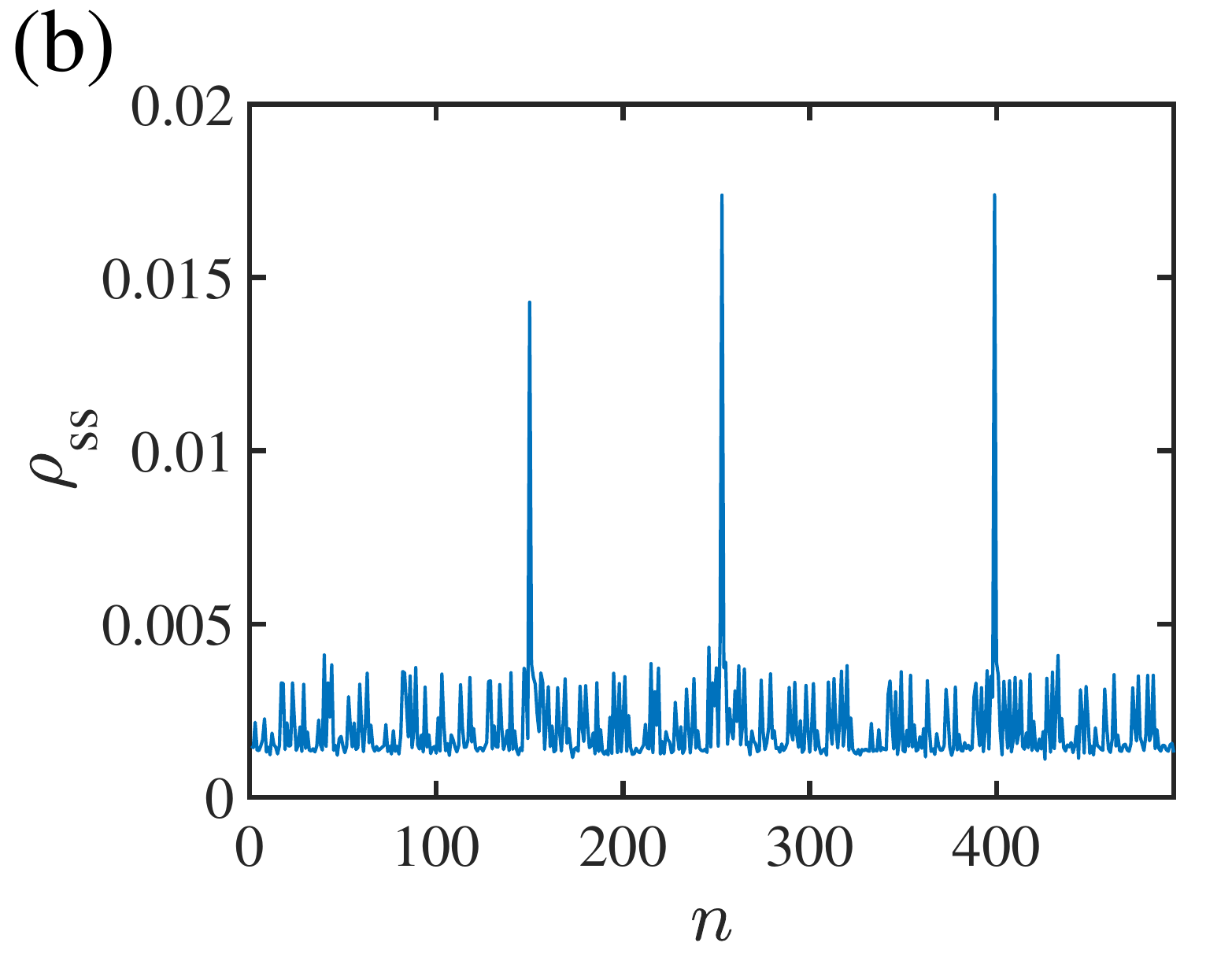} 
	\caption{(a) Fidelity dynamics for a $Z_{3}$-symmetric initial state under unitary (red) and dissipative (green) evolution. The calculation is performed for system size $L=12$ with emergent disorder strength $h/J=0.5$. The dissipative fidelity curve remains consistently higher than the unitary one, indicating that dissipation enhances memory retention. 
	(b) Diagonal elements of the steady-state density matrix versus the index of local basis. Three sharp peaks correspond to $\ket{Z_3}=\ket{100100100100}$, $\ket{Z^{\prime}_3}=\ket{010010010010}$ and $\ket{Z^{\prime\prime}_3}=\ket{001001001001}$, respectively. } 
	\label{Fig4} 
\end{figure}

The resulting dynamics are shown in Fig.~\ref{Fig4}(a). For a chain of length $L=12$, the fidelity under dissipative evolution (green) consistently lies above that of purely unitary dynamics (red), providing clear evidence that the engineered dissipation enhances the system's ability to retain memory of the $Z_{3}$ initial state. This behavior persists throughout the entire time evolution, highlighting the robustness of the dissipation-induced localization effect.

Further insight can be gained by examining the fidelity of steady-state density matrix. Fig.~\ref{Fig4}(b) plots its diagonal elements in the local basis. One can see that three pronounced peaks emerge, signifying the steady state is dominated by contributions from states with $Z_{3}$ symmetry. 


Next, we extend our investigation to an initial state exhibiting $Z_{4}$ symmetry. In this setting, the system size is chosen as $L=12$, and the jump amplitude is adjusted to $l=4$ in order to adjust the four-site periodicity of the state. The corresponding dissipation operator takes the form
\begin{align}
	O_{j} = \big( c_{\boldsymbol{j}}^\dagger + c_{\boldsymbol{j} + 4}^\dagger \big) \big( c_{\boldsymbol{j}} - c_{\boldsymbol{j} + 4} \big).
\end{align}

\begin{figure}[!ht] 
	\includegraphics[width=1.0\linewidth]{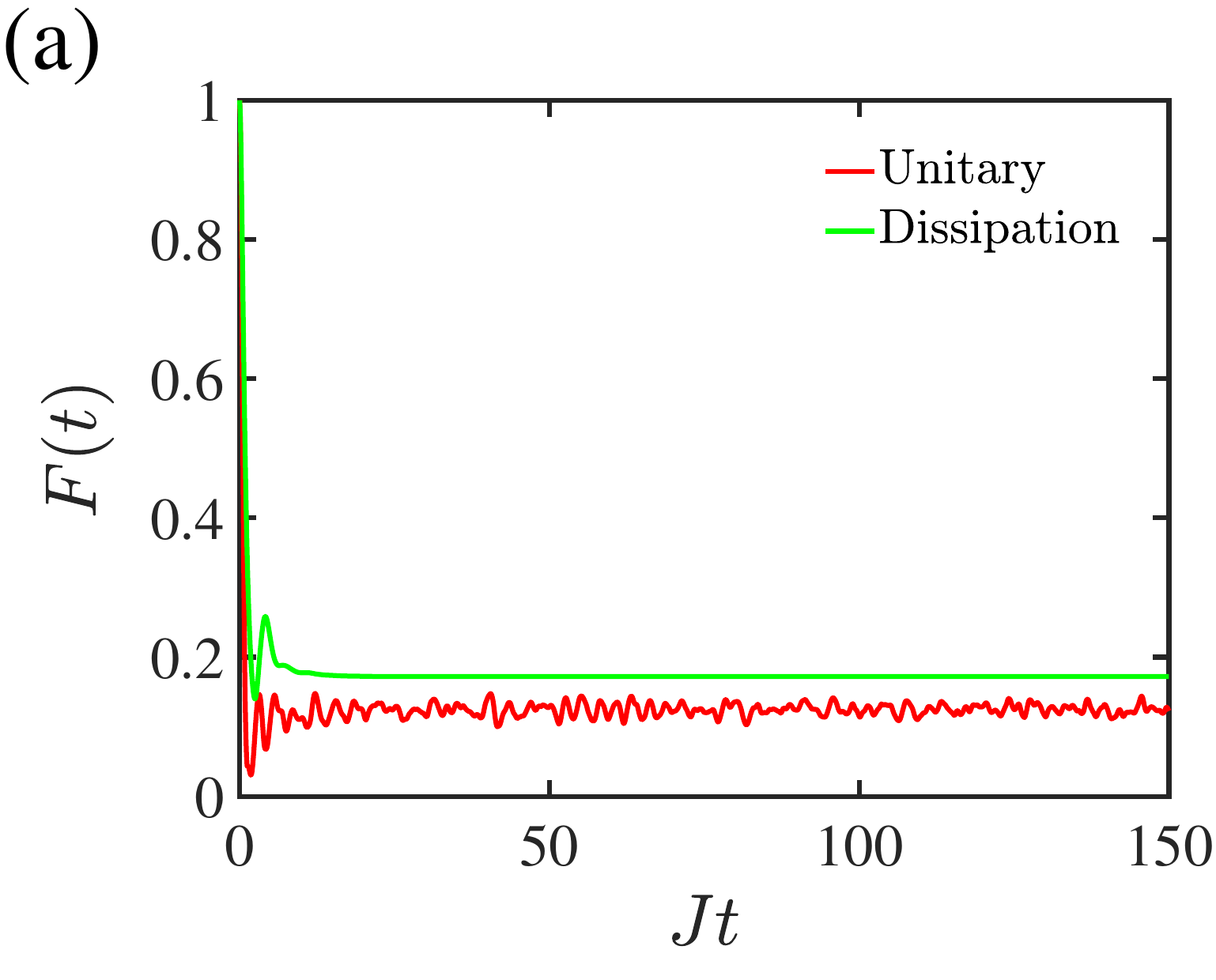} 
	\includegraphics[width=1.0\linewidth]{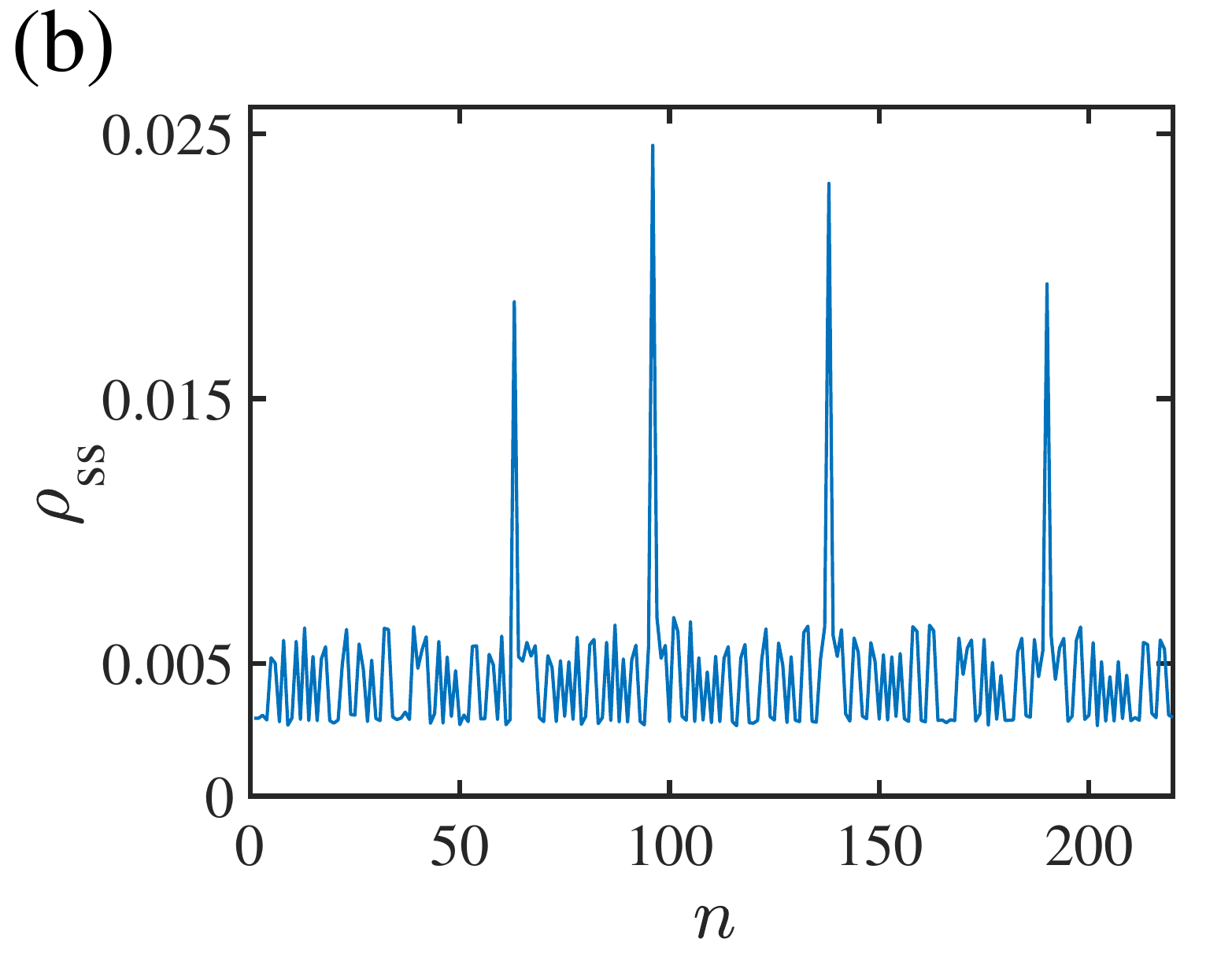} 
	\caption{(a) Fidelity dynamics for a $Z_{4}$-symmetric initial state under unitary (red) and dissipative (green) evolution. The calculation is performed for system size $L=12$ with emergent disorder strength $h/J=0.5$. The dissipative fidelity consistently lies above the unitary one, demonstrating enhanced retention of the initial-state memory. (b) Diagonal elements of the steady-state density matrix in the local basis. Four pronounced peaks are visible, corresponding to the $\ket{Z_{4}}=\ket{100010001000}$ configuration and its three translated counterparts ($Z_{4}^\prime=\ket{010001000100}$, $Z_{2}^{\prime\prime}=\ket{001000100010}$, and $Z_{2}^{\prime\prime\prime}=\ket{000100010001}$), confirming that the dissipative dynamics stabilize a $Z_{4}$-symmetric steady state.} 
	
	\label{Fig5} 
\end{figure}

The corresponding results are displayed in Fig.\ref{Fig5}. Fig.\ref{Fig5}(a) shows the fidelity dynamics under both unitary and dissipative evolution. As in the $\mathbb{Z}_2$ and $\mathbb{Z}_3$ symmetric cases, the dissipative curve (green) consistently remains above the unitary one (red), demonstrating that dissipation enhances the retention of the initial state's memory. Fig.~\ref{Fig5}(b) provides further insight by showing diagonal elements of the steady-state density matrix in the local basis. The presence of four sharp peaks reveals that the steady state is dominated by contributions from the $\ket{Z_{4}}$ configuration together with its three symmetry-related translations, $\ket{Z_{4}^\prime}$, $\ket{Z_{4}^{\prime\prime}}$, and $\ket{Z_{4}^{\prime\prime\prime}}$. This confirms that the system preferentially relaxes into a $\mathbb{Z}_4$-symmetric steady state stabilized by dissipation.

From these results, we conclude that for initial states with various $Z_{n}$ symmetries, one only needs to tune the jump amplitude $l=n$ in the dissipation operator to match the symmetry. The dissipation mechanism we propose not only avoids destroying the system's memory but in fact enhances it, leading to non-thermal steady states with symmetry-dependent structures.
These results confirm that dissipation not only preserves but also reinforces disorder-free localization beyond the $Z_{2}$ case, extending the mechanism of memory enhancement to higher-symmetry initial configurations.

%
%
%
%
%
%
%
%
%
%
%
%
%

	\section{Conclusion and outlook} 
	\label{summarize}

In this work we have investigated the role of engineered dissipation in the non-equilibrium dynamics of the $\mathbb{Z}_2$ lattice gauge model. By coupling the system to a Markovian bath with carefully designed jump operators, we demonstrated that dissipation, contrary to the conventional expectation of destroying coherence, can enhance disorder-free localization and reinforce memory of the initial state. Using quantum fidelity as a diagnostic, we showed that dissipative dynamics lead to stronger retention of symmetry-protected states compared with purely unitary evolution. Moreover, by tuning the jump amplitude to match different $\mathbb{Z}_n$ symmetries, we established a general mechanism by which dissipation can stabilize non-thermal steady states characterized by symmetry-dependent structures.

Our results highlight a counterintuitive and versatile role of quantum dissipation. Rather than erasing memory, it can be exploited as a resource to protect localization and stabilize non-ergodic dynamics in gauge-constrained systems. This opens several exciting directions for future research. On the theoretical side it would be valuable to explore the interplay between dissipation-enhanced localization and many-body interactions beyond the gauge framework, as well as to investigate robustness against different classes of environmental noise. Several particularly intriguing directions deserve to be explored. First, it would be of great interest to explore experimental implementations of such dissipation engineering in quantum simulators, including cold atoms, trapped ions, and superconducting qubits, where local control of dissipation channels has become feasible. Second, generalizing this approach to the present framework from discrete $\mathbb{Z}_{n}$ symmetries to continuous lattice gauge theories such as $U(1)$ or even non-Abelian gauge models. Understanding whether and how engineered dissipation can reinforce localization and memory in these richer settings would not only broaden the scope of non-equilibrium gauge dynamics but also bring us closer to realizing controlled non-ergodic phases in synthetic quantum matter. Finally, studying the interplay between engineered dissipation and external driving or interactions could provide new insights into stabilizing exotic non-equilibrium phases and long-lived coherent dynamics in open quantum systems.

\section*{Acknowledgements}
We thank X.-C. Zhou for helpful discussions. The work is supported by National Natural Science Foundation of China (Grant No. 12304290). LP acknowledges support from the Fundamental Research Funds for the Central Universities. \\


\begin{thebibliography}{99}
	
\bibitem{Kogut_RMP1979} J. B. Kogut, An introduction to lattice gauge theory and spin systems, \href{https://doi.org/10.1103/RevModPhys.51.659}{Rev. Mod. Phys. {\bf 51}, 659 (1979).}

\bibitem{Banerjee_PRL2012} D. Banerjee, M. Dalmonte, M. Müller, E. Rico, P. Stebler, U.-J. Wiese, and P. Zoller, Atomic quantum simulation of dynamical gauge fields coupled to fermionic matter: From string breaking to evolution after a quench, \href{https://doi.org/10.1103/PhysRevLett.109.175302}{Phys. Rev. Lett. {\bf 109}, 175302 (2012).}

\bibitem{Zohar_RPP2016} E. Zohar, J. I. Cirac, and B. Reznik, Quantum simulations of lattice gauge theories using ultracold atoms in optical lattices, \href{https://doi.org/10.1088/0034-4885/79/1/014401}{Rep. Prog. Phys. {\bf 79}, 014401 (2016).}

\bibitem{Martinez_Nature2016} E. A. Martinez, C. A. Muschik, P. Schindler, D. Nigg, A. Erhard, M. Heyl, P. Hauke, M. Dalmonte, T. Monz, P. Zoller, and R. Blatt, Real-time dynamics of lattice gauge theories with a few-qubit quantum computer, \href{https://doi.org/10.1038/nature18318}{Nature {\bf 534}, 516 (2016).}


\bibitem{Gauge_exp1} B. Yang, H. Sun, R. Ott, H. Y. Wang, T. V. Zache, J. C. Halimeh, Z. S. Yuan, P. Hauke, J. W. Pan, Observation of gauge invariance in a 71-site Bose-Hubbard quantum simulator. \href{https://doi.org/10.1038/s41586-020-2910-8}{Nature {\bf 587}, 392 (2020).}

\bibitem{Gauge_exp2} Z. Y. Zhou, G. X. Su, J. C. Halimeh, R. Ott, H. Sun, P. Hauke, B. Yang, Z. S. Yuan, J. Berges, J. W. Pan, Thermalization dynamics of a gauge theory on a quantum simulator, \href{DOI: 10.1126/science.abl6277}{Science {\bf 377}, 311 (2022).}

\bibitem{Gauge_exp3} H. Y. Wang, W. Y. Zhang, Z. Yao, Y. Liu, Z. H. Zhu, Y. G. Zheng, X. K. Wang, H. Zhai, Z. S. Yuan, and J. W. Pan, Interrelated thermalization and quantum criticality in a lattice gauge simulator. \href{https://doi.org/10.1103/PhysRevLett.131.050401}{Phys. Rev. Lett. {\bf 131}, 050401 (2023).}

\bibitem{Gauge_exp4} J.-Y. Desaules, G.-X. Su, I. P. McCulloch, B. Yang, Z. Papi{\'c}, and J. C. Halimeh, Ergodicity Breaking Under Confinement in Cold-Atom Quantum,  Simulators, \href{https://doi.org/10.22331/q-2024-02-29-1274}{Quantum {\bf 8}, 1274 (2024).}

\bibitem{Gauge_exp5} W.-Y. Zhang, Y. Liu, Y. Cheng, M.-G. He, H.-Y. Wang, T.-Y. Wang, Z.-H. Zhu, G.-X. Su, Z.-Y. Zhou, Y.-G. Zheng, H. Sun, B. Yang, P. Hauke, W. Zheng, J. C. Halimeh, Z. S. Yuan, and J. W. Pan, 
Observation of microscopic confinement dynamics by a tunable topological $\theta$-angle, \href{https://doi.org/10.1038/s41567-024-02702-x}{Nat. Phys. {\bf 21}, 155 (2025).}

\bibitem{Gauge_exp6} J. J. Osborne, I. P. McCulloch, B. Yang, P. Hauke, and J. C. Halimeh,  Large-scale 2+1D $U(1)$ gauge theory with dynamical matter in a cold-atom quantum simulator, \href{https://doi.org/10.1038/s42005-025-02144-8}{Commun. Phys. {\bf 8} 273 (2025). }

\bibitem{Gauge_exp7} J. C. Halimeh, M. Aidelsburger, F. Grusdt, P. Hauke, and B. Yang , Cold-atom quantum simulators of gauge theories, \href{https://doi.org/10.1038/s41567-024-02721-8}{Nat. Phys. {\bf 21} 25 (2025).}

\bibitem{Gauge_exp8} J. C. Halimeh, N. Mueller, J. Knolle, Z. Papi{\'c}, Zohreh Davoudi, Quantum simulation of out-of-equilibrium dynamics in gauge theories, \href{ 	https://doi.org/10.48550/arXiv.2509.03586}{arXiv:2509. 03586 (2025).}



\bibitem{Z2_theory1} A. Smith, J. Knolle, D. L. Kovrizhin, and R. Moessner, Disorder-Free Localization, \href{https://doi.org/10.1103/PhysRevLett.118.266601}{Phys. Rev. Lett. {\bf 118}, 266601 (2017).}

\bibitem{Z2_theory2} A. Smith, J. Knolle, R. Moessner, and D. L. Kovrizhin, Dynamical localization in $\mathbb{Z}_2$
lattice gauge theories
 \href{https://doi.org/10.1103/PhysRevB.97.245137}{Phys. Rev. B {\bf 97}, 245137 (2018).}

\bibitem{Anderson_PR1958} P. W. Anderson, Absence of diffusion in certain random lattices, \href{https://doi.org/10.1103/PhysRev.109.1492}{Phys. Rev. {\bf 109}, 1492 (1958).}

\bibitem{Nandkishore_ARCMP2015} R. Nandkishore and D. A. Huse, Many-body localization and thermalization in quantum statistical mechanics, \href{https://doi.org/10.1146/annurev-conmatphys-031214-014726}{Annu. Rev. Condens. Matter Phys. {\bf 6}, 15 (2015).}

\bibitem{AndersonTrans2008} F. Evers and A. D. Mirlin, Anderson transitions
\href{https://journals.aps.org/rmp/abstract/10.1103/RevModPhys.80.1355}{Rev. Mod. Phys. {\bf 80}, 1355 (2008).}	

\bibitem{Exp1} G. Barontini, R. Labouvie, F. Stubenrauch, A. Vogler, V. Guarrera, and H. Ott,
Controlling the Dynamics of an Open Many-Body Quantum System with Localized Dissipation,
\href{https://journals.aps.org/prl/abstract/10.1103/PhysRevLett.110.035302}{Phys. Rev. Lett. {\bf 110}, 035302 (2013).}

\bibitem{Exp2} Y. S. Patil, S. Chakram, and M. Vengalattore,
Measurement-Induced Localization of an Ultracold Lattice Gas,
\href{https://journals.aps.org/prl/abstract/10.1103/PhysRevLett.115.140402}{Phys. Rev. Lett. {\bf 115}, 140402 (2015).}

\bibitem{Exp3} R. Labouvie, B. Santra, S. Heun, and H. Ott,
Bistability in a Driven-Dissipative Superfluid,
\href{https://journals.aps.org/prl/abstract/10.1103/PhysRevLett.116.235302}{Phys. Rev. Lett. {\bf 116}, 235302 (2016).}

\bibitem{Exp4} H. P. L{\"u}schen, P. Bordia, S. S. Hodgman, M. Schreiber, S. Sarkar, A. J. Daley, M. H. Fischer, E. Altman, I. Bloch, and U. Schneider,
Signatures of Many-Body Localization in a Controlled Open Quantum System,
\href{https://journals.aps.org/prx/abstract/10.1103/PhysRevX.7.011034}{Phys. Rev. X {\bf 7}, 011034 (2017).}

\bibitem{Exp5} T. Tomita, S. Nakajima, Y. Takasu, and Y. Takahashi,
Dissipative Bose-Hubbard system with intrinsic two-body loss,
\href{https://journals.aps.org/pra/abstract/10.1103/PhysRevA.99.031601}{Phys. Rev. A {\bf 99}, 031601(R) (2019).}
\bibitem{Exp6} R. Bouganne, M. B. Aguilera, A. Ghermaoui, and F. Gerbier,
Anomalous decay of coherence in a dissipative many-body system,
\href{https://www.nature.com/articles/s41567-019-0678-2}{Nat. Phys. 16, 21 (2020).}

\bibitem{Exp7} T. Tomita, S. Nakajima, I. Danshita, Y. Takasu, and
Y. Takahashi,
Observation of the Mott insulator to superfluid crossover of a driven-dissipative Bose-Hubbard system,
\href{https://www.science.org/doi/full/10.1126/sciadv.1701513}{Sci. Adv. {\bf 3}, e1701513 (2017).}
\bibitem{Exp8} K. Sponselee, L. Freystatzky, B. Abeln, M. Diem, B. Hundt,
Dynamics of ultracold quantum gases in the dissipative Fermi–Hubbard model,
A. Kochanke, T. Ponath, B. Santra, L. Mathey, K. Sengstock, and C. Becker, \href{https://iopscience.iop.org/article/10.1088/2058-9565/aadccd/meta}{Quantum Sci. Technol. {\bf 4}, 014002 (2018).}

\bibitem{Exp9} Y. Takasu, T. Yagami, Y. Ashida, R. Hamazaki, Y. Kuno, and
Y. Takahashi,
PT-symmetric non-Hermitian quantum many-body system using ultracold atoms in an optical lattice with controlled dissipation,
\href{https://academic.oup.com/ptep/article/2020/12/12A110/5905047?login=true}{Prog. Theor. Exp. Phys. {\bf 2020}, 12A110 (2020).}

\bibitem{Exp10} B. Yan, S. A. Moses, B. Gadway, J. P. Covey, K. R. Hazzard, A. M. Rey, D. S. Jin, and J. Ye,
Observation of dipolar spin-exchange interactions with lattice-confined polar molecules,
\href{https://www.nature.com/articles/nature12483}{Nature (London) 501, 521 (2013).}

\bibitem{Exp11} F. Sch{\"a}fer, T. Fukuhara, S. Sugawa, Y. Takasu, and Y. Takahashi,
Tools for quantum simulation with ultracold atoms in optical lattices,
\href{https://www.nature.com/articles/s42254-020-0195-3}{ Nat. Rev. Phys. {\bf 2}, 411 (2020).}  

\bibitem{Exp12} R. Bouganne, M. B. Aguilera, A. Ghermaoui, and F. Gerbier, Anomalous decay of coherence in a dissipative many-body system, \href{https://doi.org/10.1038/s41567-019-0678-2}{Nat. Phys. {\bf 16}, 21 (2020). }

\bibitem{Exp13} Y. Zhao, Y. Tian, J. Ye, Y. Wu, Z. Zhao, Z. Chi, T. Tian, H. Yao, J. Hu, Y. Chen, and W. Chen, Universal dissipative dynamics in strongly correlated quantum gases, \href{https://doi.org/10.1038/s41567-025-02800-4}{Nat. Phys. {\bf 21}, 530 (2025).}



\bibitem{OpenMB0} K. Yamamoto, M. Nakagawa, K. Adachi, K. Takasan, M. Ueda, and N. Kawakami,
Theory of Non-Hermitian Fermionic Superfluidity with a Complex-Valued Interaction,
\href{https://journals.aps.org/prl/abstract/10.1103/PhysRevLett.123.123601}{Phys. Rev. Lett. {\bf 123}, 123601 (2019).}


\bibitem{OpenMB1} L. Zhou and X. Cui,
Enhanced Fermion Pairing and Superfluidity by an Imaginary Magnetic Field,
\href{https://www.sciencedirect.com/science/article/pii/S2589004219300975}{iScience {\bf 14}, 257 (2019).}

\bibitem{OpenMB2} L. Pan, X. Chen, Y. Chen, H. Zhai,
Non-Hermitian linear response theory,
\href{https://www.nature.com/articles/s41567-020-0889-6}{Nat. Phys. {\bf 16}, 767 (2020).}

\bibitem{OpenMB3} Z. Cai and T. Barthel,
Algebraic versus Exponential Decoherence in Dissipative Many-Particle Systems,
\href{https://journals.aps.org/prl/abstract/10.1103/PhysRevLett.111.150403}{Phys. Rev. Lett. {\bf 111}, 150403 (2013).}

\bibitem{OpenMB4} K. Yamamoto, M. Nakagawa, N. Tsuji, M. Ueda, and N. Kawakami,
Collective Excitations and Nonequilibrium Phase Transition in Dissipative Fermionic Superfluids,
\href{https://journals.aps.org/prl/abstract/10.1103/PhysRevLett.127.055301}{Phys. Rev. Lett. {\bf 127}, 055301 (2021).}	


\bibitem{OpenMB5} K. L. Zhang and Z. Song,
Quantum Phase Transition in a Quantum Ising Chain at Nonzero Temperatures,
\href{https://journals.aps.org/prl/abstract/10.1103/PhysRevLett.126.116401}{Phys. Rev. Lett. {\bf 126}, 116401 (2021).}


\bibitem{OpenMB6} {\'A}. B{\'a}csi, C. P. Moca, and B. D{\'o}ra,
Dissipation-Induced Luttinger Liquid Correlations in a One-Dimensional Fermi Gas,
\href{https://journals.aps.org/prl/abstract/10.1103/PhysRevLett.124.136401}{Phys. Rev. Lett. 124, 136401 (2020).}

\bibitem{OpenMB7} L. Pan, X. Wang, X. Cui, and S. Chen, 
Interaction-induced dynamical $\mathscr{PT}$
-symmetry breaking in dissipative Fermi-Hubbard models,
\href{https://journals.aps.org/pra/abstract/10.1103/PhysRevA.102.023306}{Phys. Rev. A {\bf 102}, 023306 (2020).}

\bibitem{OpenMB8} X. Z. Zhang and Z. Song,
Dynamical preparation of a steady off-diagonal long-range order state in the Hubbard model with a local non-Hermitian impurity,
\href{https://journals.aps.org/prb/abstract/10.1103/PhysRevB.102.174303}{Phys. Rev. B {\bf 102}, 174303 (2020).}

\bibitem{OpenMB9} K. Yang, S. C. Morampudi, and E. J. Bergholtz,
Exceptional Spin Liquids from Couplings to the Environment,
\href{https://journals.aps.org/prl/abstract/10.1103/PhysRevLett.126.077201}{Phys. Rev. Lett. {\bf 126}, 0772012 (2021).}


\bibitem{OpenMB10} M. Nakagawa, N. Kawakami, and M. Ueda,
Exact Liouvillian Spectrum of a One-Dimensional Dissipative Hubbard Model,
\href{https://journals.aps.org/prl/abstract/10.1103/PhysRevLett.126.110404}{Phys. Rev. Lett. 126, 110404 (2021).}

\bibitem{OpenMB11} M. Nakagawa, N. Tsuji, N. Kawakami, and M. Ueda, 
Dynamical Sign Reversal of Magnetic Correlations in Dissipative Hubbard Models,
\href{https://journals.aps.org/prl/abstract/10.1103/PhysRevLett.124.147203}{Phys. Rev. Lett. 124, 147203 (2020).}

\bibitem{OpenMB12} L. S{\'a}, P. Ribeiro, and T. Prosen,
Complex Spacing Ratios: A Signature of Dissipative Quantum Chaos,
\href{https://journals.aps.org/prx/abstract/10.1103/PhysRevX.10.021019}{Phys. Rev. X {\bf 10}, 021019 (2020).}

\bibitem{OpenMB13} J. Li, T. Prosen, and A. Chan,
Spectral Statistics of Non-Hermitian Matrices and Dissipative Quantum Chaos,
\href{https://journals.aps.org/prl/abstract/10.1103/PhysRevLett.127.170602}{Phys. Rev. Lett. {\bf 127}, 170602 (2021).} 


\bibitem{OpenMB14} T. Mori,
Liouvillian-gap analysis of open quantum many-body systems in the weak dissipation limit,
\href{https://arxiv.org/abs/2311.10304}{arXiv:2311.10304 (2023).}

\bibitem{OpenMB15} C.-Z. Lu, X. Deng, S.-P. Kou, G. Sun,
Unconventional many-body phase transitions in a non-Hermitian Ising chain,
\href{https://doi.org/10.1103/PhysRevB.110.014441}{Phys. Rev. B {\bf 110}, 014441 (2024).}


\bibitem{OpenMB16} Non-Hermitian skin effect in a one-dimensional interacting Bose gas,
L. Mao, Y. Hao, and L. Pan,
\href{https://journals.aps.org/pra/abstract/10.1103/PhysRevA.107.043315}{Phys. Rev. A {\bf 107}, 043315 (2023).}

\bibitem{OpenMB17} L. Mao, X. Yang, M.-J. Tao, H. Hu, and L. Pan, Liouvillian skin effect in a one-dimensional open many-body quantum system with generalized boundary conditions,
\href{https://journals.aps.org/prb/abstract/10.1103/PhysRevB.110.045440}{Phys. Rev. B {\bf 110}, 045440 (2024).}




\bibitem{OpenMBL1} R. Hamazaki, K. Kawabata, and M. Ueda,
Non-Hermitian Many-Body Localization,
\href{https://journals.aps.org/prl/abstract/10.1103/PhysRevLett.123.090603}{Phys. Rev. Lett. {\bf 123}, 090603 (2019).} 

\bibitem{OpenMBL2} L.-J. Zhai, S. Yin, and G.-Y. Huang,
Many-body localization in a non-Hermitian quasiperiodic system,
\href{https://journals.aps.org/prb/abstract/10.1103/PhysRevB.102.064206}{Phys. Rev. B {\bf 102}, 064206 (2020).}


\bibitem{OpenMBL3} K. Suthar, Y.-C. Wang, Y.-P. Huang, H.-H. Jen, and J.-S. You, Non-Hermitian many-body localization with open boundaries,
\href{https://doi.org/10.1103/PhysRevB.106.064208}{Phys. Rev. B {\bf 106}, 0642085 (2022).}


\bibitem{OpenMBL4} F. Roccati, F. Balducci, R. Shir, and A. Chenu, Diagnosing non-Hermitian many-body localization and quantum chaos via singular value decomposition,  \href{https://journals.aps.org/prb/abstract/10.1103/PhysRevB.109.L140201}{Phys. Rev. B  {\bf 109}, L140201 (2024).}






\bibitem{OpenDisorder1}  V. Balachandran, S. R. Clark, J. Goold, and D. Poletti, Energy Current Rectification and Mobility Edges, \href{https://journals.aps.org/prl/abstract/10.1103/PhysRevLett.123.020603}{Phys. Rev. Lett. {\bf 123}, 020603 (2019).}

\bibitem{OpenDisorder2} C. Chiaracane, M. T. Mitchison, A. Purkayastha,
G. Haack, and J. Goold, Quasiperiodic quantum heat engines with a mobility edge, \href{https://journals.aps.org/prresearch/abstract/10.1103/PhysRevResearch.2.013093}{Phys. Rev. Research {\bf 2}, 013093 (2020).}

\bibitem{OpenDisorder3} M. Balasubrahmaniyam, S. Mondal, and S. Mujumdar, Necklace-State-Mediated Anomalous Enhancement of Transport in Anderson-Localized non-Hermitian Hybrid Systems, \href{https://journals.aps.org/prl/abstract/10.1103/PhysRevLett.124.123901}{Phys. Rev. Lett. {\bf 124}, 123901 (2020).}

\bibitem{OpenDisorder4} S. Weidemann, M. Kremer, S. Longhi, and A. Szameit, Coexistence of dynamical delocalization and spectral localization through stochastic dissipation, \href{https://www.nature.com/articles/s41566-021-00823-w}{Nat. Photon. {\bf 15}, 576 (2021).}


\bibitem{OpenDisorder5} C. Chiaracane, A. Purkayastha, M. T. Mitchison, and J. Goold, Dephasing-enhanced performance in quasiperiodic thermal machines,  \href{https://journals.aps.org/prb/abstract/10.1103/PhysRevB.105.134203}{Phys. Rev. B {\bf 105}, 134203 (2022).}

\bibitem{OpenDisorder6} S. Longhi, Anderson Localization in Dissipative Lattices, \href{https://doi.org/10.1002/andp.202200658}{Ann. Phys. {\bf 535}, 2200658 (2023).}

\bibitem{OpenDisorder7} S. Longhi, Dephasing-Induced Mobility Edges in Quasicrystals, 
\href{https://journals.aps.org/prl/abstract/10.1103/PhysRevLett.132.236301}{Phys. Rev. Lett. 132, 236301 (2024).}

\bibitem{OpenDisorder8} X.-P. Jiang, X. Yang, Y. Hu, L. Pan, Dissipation induced ergodic-nonergodic transitions in finite-height mosaic Wannier-Stark lattices, \href{https://arxiv.org/abs/2407.17301}{ arXiv:2407.17301 (2024). }

\bibitem{OpenDisorder9} C. Wang, and X. R. Wang, Anderson localization transitions in disordered non-Hermitian systems with exceptional points \href{https://journals.aps.org/prb/abstract/10.1103/PhysRevB.107.024202}{Phys. Rev. B \textbf{107}, 024202 (2023).}



\bibitem{OpenDisorder10} L.-J. Zhai, S. Yin, and G.-Y. Huang, Many-body localization in a non-Hermitian quasiperiodic system, \href{https://journals.aps.org/prb/abstract/10.1103/PhysRevB.102.064206}{Phys. Rev. B \textbf{102}, 064206 (2020).}

\bibitem{OpenDisorder11} Y.-C. Wang, K. Suthar, H. H. Jen, Y.-T. Hsu, and J.-S. You, Non-Hermitian skin effects on thermal and many-body localized phases, \href{https://journals.aps.org/prb/abstract/10.1103/PhysRevB.107.L220205}{Phys. Rev. B \textbf{107}, L220205 (2023).}

\bibitem{OpenDisorder12} Y. Huang and B. I. Shklovskii, Spectral rigidity of non-Hermitian symmetric random matrices near the Anderson transition, \href{https://journals.aps.org/prb/abstract/10.1103/PhysRevB.102.064212}{Phys. Rev. B \textbf{102}, 064212 (2020).}

\bibitem{OpenDisorder13} Y. Huang and B. I. Shklovskii, Anderson transition in three-dimensional systems with non-Hermitian disorder, \href{https://journals.aps.org/prb/abstract/10.1103/PhysRevB.101.014204}{Phys. Rev. B \textbf{101}, 014204 (2020).} 

\bibitem{OpenDisorder14} Y. Peng, C. Yang, and Y. Wang, Manipulating the relaxation time of boundary-dissipative systems through bond dissipation, \href{https://doi.org/10.1103/PhysRevB.110.104305}{Phys. Rev. B \textbf{110}, 104305 (2024).}


\bibitem{NH_Z2} J.-Q. Cheng, S. Yin, and D.-X. Yao, Dynamical localization transition in the non-Hermitian lattice gauge theory, \href{https://doi.org/10.1038/s42005-024-01544-6}{Commun. Phys. {\bf 7}, 58 (2024).}

\bibitem{OpenDisorder_transport1} A. M. Lacerda, J. Goold, and G. T. Landi, Dephasing enhanced transport in boundary-driven quasiperiodic chains, \href{https://journals.aps.org/prb/abstract/10.1103/PhysRevB.104.174203}{Phys. Rev. B {\bf 104}, 174203 (2021).}

\bibitem{OpenDisorder_transport2} D. Dwiputra and F. P. Zen, Environment-assisted quantum transport and mobility edges, \href{https://journals.aps.org/pra/abstract/10.1103/PhysRevA.104.022205}{Phys. Rev. A {\bf 104}, 022205 (2021).}


\bibitem{OpenDisorder_transport3} M. Saha, B. P. Venkatesh, and B. K. Agarwalla, Quantum transport in quasiperiodic lattice systems in the presence of B\"{u}ttiker probes, \href{https://journals.aps.org/prb/abstract/10.1103/PhysRevB.105.224204}{Phys. Rev. B {\bf 105}, 224204 (2022).}


\bibitem{Yusipov17} I. Yusipov, T. Laptyeva, S. Denisov, and M. Ivanchenko, Localization in Open Quantum Systems, \href{https://journals.aps.org/prl/abstract/10.1103/PhysRevLett.118.070402}{Phys. Rev. Lett. {\bf 118}, 070402 (2017).}

\bibitem{WYC_PRL} Y. Liu, Z. Wang, C. Yang, J. Jie, and Y. Wang, Dissipation-Induced Extended-Localized Transition, \href{https://journals.aps.org/prl/abstract/10.1103/PhysRevLett.132.216301}{Phys. Rev. Lett. {\bf 132}, 216301 (2024).}

\bibitem{Jiang_3D} X. Yang, X.-P. Jiang, Z. Wei, Y. Wang, and L. Pan, Dissipation-induced transition between delocalization and localization in the three-dimensional Anderson model, \href{https://doi.org/10.1103/PhysRevB.111.134203}{Phys. Rev. B {\bf 111}, 134203 (2025).}

\bibitem{Yusipov18} I. Vakulchyk, I. Yusipov, M. Ivanchenko, S. Flach, and S. Denisov, Signatures of many-body localization in steady states of open quantum systems, \href{https://journals.aps.org/prb/abstract/10.1103/PhysRevB.98.020202}{Phys. Rev. B {\bf 98}, 020202(R) (2018).}

\bibitem{WYC_MBL} Y. Hu, C. Yang, and Y. Wang, Can dissipation induce a transition between many-body localized and thermal states? 
\href{https://arxiv.org/abs/2407.13655}{arXiv:2407.13655 (2024).}

\bibitem{Diss_scar1} X.-P. Jiang, M. Xu, X. Yang, H. Hou, Y. Wang, and L. Pan, Robustness of quantum many-body scars in the presence of Markovian bath, \href{https://arxiv.org/abs/2501.00886}{ 	arXiv:2501.00886 (2025).}

\bibitem{Diss_scar2}  H.-R. Wang, D. Yuan, S.-Y. Zhang, Z. Wang, D.-L. Deng, and L.-M. Duan, Embedding Quantum Many-Body Scars into Decoherence-Free Subspaces, \href{https://journals.aps.org/prl/abstract/10.1103/PhysRevLett.132.150401}{Phys. Rev. Lett. {\bf 132}, 150401 (2024).}

\bibitem{Diss_scar3}  R. Shen, F. Qin, J.-Y. Desaules, Z. Papi{\'c}, and C. H. Lee, Enhanced Many-Body Quantum Scars from the Non-Hermitian Fock Skin Effect, \href{https://journals.aps.org/prl/abstract/10.1103/PhysRevLett.133.216601}{Phys. Rev. Lett. {\bf 133}, 216601 (2024).}

\bibitem{duality1}  H. A. Kramers and G. H. Wannier, \href{https://doi.org/10.1103/PhysRev.60.252}{Phys. Rev. {\bf 60}, 252 (1941).}
\bibitem{duality2}  E. Fradkin and L. Susskind, \href{https://doi.org/10.1103/PhysRevD.17.2637}{Phys. Rev. D {\bf 17}, 2637 (1978).}

\bibitem{fidelity} P. Zanardi, H. T. Quan, X. Wang, and C. P. Sun,
Mixed-state fidelity and quantum criticality at finite temperature,
\href{https://doi.org/10.1103/PhysRevA.75.032109}{Phys. Rev. A {\bf 75}, 032109 (2007).}

\bibitem{Moy1999}
G. M. Moy, J. J. Hope, and C. M. Savage,
Born and Markov approximations for atom lasers,
\href{https://doi.org/10.1103/PhysRevA.59.667}{Phys. Rev. A \textbf{59}, 667 (1999).}

\bibitem{Breuer2002}
H.-P. Breuer and F. Petruccione,
\textit{The Theory of Open Quantum Systems},
(Oxford University Press, Oxford, 2002).

\bibitem{Lindblad1} G. Lindblad, On the generators of quantum dynamical semigroups,
\href{https://link.springer.com/article/10.1007/BF01608499}{Commun. Math. Phys. {\bf 119}, 48 (1976).}

\bibitem{Lindblad2} V. Gorini, A. Kossakowski, and E. C. Sudarsahan,
Completely positive dynamical semigroups of N‐level systems,
\href{https://pubs.aip.org/aip/jmp/article/17/5/821/225427/Completely-positive-dynamical-semigroups-of-N}{J. Math. Phys. {\bf 17}, 821 (1976).}

%

\bibitem{Jump1} S. Diehl, A. Micheli, A. Kantian, B. Kraus, H. P. Büchler, and P. Zoller, Quantum states and phases in driven open quantum systems with cold atoms,  \href{https://www.nature.com/articles/nphys1073}{Nat. Phys. {\bf 4}, 878 (2008).}

\bibitem{Jump2} B. Kraus, H. P. B{\"u}chler, S. Diehl, A. Kantian, A. Micheli, and P. Zoller, Preparation of entangled states by quantum Markov processes,  \href{https://journals.aps.org/pra/abstract/10.1103/PhysRevA.78.042307}{Phys. Rev. A {\bf 78}, 042307 (2008).}

\bibitem{BHchain} D. Marcos, A. Tomadin, S. Diehl, and P. Rabl,
Photon condensation in circuit quantum electrodynamics by engineered dissipation, \href{https://iopscience.iop.org/article/10.1088/1367-2630/14/5/055005}{New J. Phys. {\bf 14}, 055005 (2012).}







%

%


	

\end{thebibliography}
\end{document}